\documentstyle[twocolumn,aps,url]{revtex}						
\newcommand{\ssub}{{\raisebox{1.5pt}{\tiny $-$} }}
\newcommand{\sadd}{{\raisebox{1.5pt}{\tiny $+$} }}
\newcommand{\tb}{{\!\!\;}} 

\begin{document}
	\wideabs{ 
	\title{Path ensembles averages in systems driven far-from-equilibrium}
	\author{Gavin E. Crooks}
	\address{Department of Chemistry, University of California at Berkeley, Berkeley, 
	CA 94720,\\ gavinc@garnet.berkeley.edu}


	\maketitle
	
	\begin{abstract}
 	  The Kawasaki nonlinear response relation, the transient 
  		fluctuation theorem, and the Jarzynski nonequilibrium work 
  		relation are all expressions that describe the behavior of a 
  		system that has been driven from equilibrium by an external 
  		perturbation. In contrast to linear response theory, these expressions are 
  		exact no matter the strength of the perturbation, or how 
  		far the system has been driven away from equilibrium. In this paper 
  		I show that these three relations (and several other closely related 
  		results) can all be considered special cases of 
  		a single theorem. This 
  		expression is explicitly derived for discrete time and space Markovian 
  		dynamics, with the additional 
  		assumptions that the single time step dynamics preserve the 
  		appropriate equilibrium ensemble, and that the energy of the 
  		system remains finite.
	\end{abstract}

} 


\section{Introduction}

If a system is gently driven from equilibrium by a small 
time-dependent perturbation, then the response of the system to the 
perturbation can be described by linear response theory. On the other 
hand, if the system is driven far-from-equilibrium by a large perturbation 
then linear response, and other near-equilibrium approximations, are 
generally not applicable.  However, there are a few relations that describe the 
statistical dynamics of driven systems which are valid even if the 
system is driven far-from-equilibrium. These include the 
Jarzynski nonequilibrium work 
relation\cite{Jarzynski97a,Jarzynski97b,Jarzynski98b,Crooks98a}, which 
gives equilibrium free energy differences in terms of nonequilibrium 
measurements of the work required to switch from one ensemble to 
another; the Kawasaki 
relation\cite{YamadaKawasaki67,MorrissEvans85,EvansMorriss90,EvansSearles95,PetravicEvans98}, 
which specifies the nonlinear response of a classical system to an 
arbitrarily large perturbation; and a group of relations that can be 
collectively called ``entropy production fluctuation 
theorems''\cite{ECM93,EvansSearles94,GC95a,GC95b,EvansSearles96,%
Gallavotti96,Cohen97,Gallavotti98,Kurchan98,Ruelle98,Lebowitz98,Maes98,Crooks99b,GC99,Evans99a,MaesRedigMoffaert99,Jarzynski99b}. 
I will specifically consider the transient fluctuation theorem of 
Evans and Searles\cite{EvansSearles94,EvansSearles96} which deals with 
entropy production of driven systems that are initially in 
equilibrium.  The Gallavotti-Cohen\cite{GC95a,GC95b} 
fluctuation theorem addresses entropy production in nonequilibrium 
steady-states, and will not be considered in this paper.
 
The relations listed above have been derived for a wide range of deterministic 
and stochastic dynamics.  However, the various expressions and 
applicable dynamics have several commonalities: the system starts in 
thermal equilibrium, it is driven from that equilibrium by an external 
perturbation, the energy of the system is finite, the dynamics are 
Markovian, and if the system is unperturbed then the dynamics preserve 
the equilibrium ensemble.  In this paper, it will be shown that these 
conditions are sufficient to derive the far-from-equilibrium 
expressions mentioned above.  Indeed they can all be considered 
special cases of a single theorem: 
\begin{equation}
	\big\langle  {\mathcal F}\big\rangle_{\!\mathrm F}
		=\big\langle \widehat{\mathcal F}\,e^{-\beta {\mathcal W}_{\!\mathrm 
		d}}  \big\rangle_{\!\mathrm R} 
\label{EqPathEnsembleAvg}.
\end{equation}

\noindent Here, $\langle {\mathcal F} \rangle_{\mathrm F}$ indicates 
the average of the path function $\mathcal F$.  Path functions (such 
as the heat and work) are functionals of the trajectory that the 
system takes through phase-space.  An average of a path function is 
implicitly an average over a suitably defined ensemble of paths.  In 
this paper, the path ensemble is defined by the initial thermal 
equilibrium and the process by which the system is subsequently 
perturbed from that equilibrium.  The left side of the above relation 
is simply $\mathcal F$ averaged over the ensemble of paths generated 
by this process.  We arbitrarily label this the forward process 
(subscript `F').
 	
For every such process that perturbs the system from equilibrium we 
can imagine a corresponding reverse perturbation (subscript `R').  We 
shall construct this process by insisting that it too starts from 
equilibrium, and by considering a formal time reversal of the 
dynamics.  The right side of Eq.~(\ref{EqPathEnsembleAvg}) is 
$\mathcal \widehat{F}$, the time reversal of $\mathcal F$, averaged 
over the reverse process, and weighted by the exponential of $\beta 
{\mathcal W}_{\mathrm d}$.  Here $\beta =1/k_{\!\scriptscriptstyle 
\mathrm B}T$, $T$ is the temperature of the heat bath, 
$k_{\!\scriptscriptstyle \mathrm B}$ is Boltzmann's constant, and 
${\mathcal W}_{\mathrm d}$ is the dissipative work.  The dissipative 
work is a path function and is defined as ${\mathcal W}_{\mathrm d} = 
{\mathcal W} - W_{\mathrm r}$, where ${\mathcal W}$ is the total work 
done on the system by the external perturbation and $W_{\mathrm r}$ 
is the reversible work, the minimum average amount of work required to 
perturb the system from its initial to its final ensemble.

In summary, Eq.~(\ref{EqPathEnsembleAvg}) states that an average taken 
over an ensemble of paths, which is generated by perturbing a system 
that is initially in equilibrium, can be equated with the average of 
another, closely related quantity, averaged over a path ensemble 
generated by the reverse process.  This relation is valid for systems 
driven arbitrarily far-from-equilibrium, and several other 
far-from-equilibrium relations can be derived from it.  It is 
sufficient that the dynamics are Markovian, preserve the equilibrium 
ensemble, and that the energy of the system is finite.  In the next 
section I derive from these conditions that such a system is 
microscopically reversible, Eq.~(\ref{EqMR}), in a sense that will be 
made precise.  (This derivation is somewhat more general than that 
given previously\cite{Crooks98a}.)  The path ensemble average, 
Eq.~(\ref{EqPathEnsembleAvg}), is an almost trivial identity given 
that the dynamics satisfy this condition.  This derivation is given is 
Sec.~(3), and various special cases are considered.

\section{Microscopic reversibility of driven systems}

Let us consider a classical system which can exchange energy with a 
constant temperature heat bath, and which has a finite set of states, 
$x \in\{1,2,3,\cdots,N\}$.  The energies of the states of the system 
are given by the vector $\mathbf E$.  If these state energies do not 
vary with time then the stationary probability distribution, $\pi$, is 
given by the canonical ensemble of equilibrium statistical mechanics; 
	\begin{eqnarray}
		\rho(x|\beta,{\mathbf E})=\pi_{x} & = & 
		\frac{e^{-\beta E_{x}}}
				{\displaystyle \sum_{x} e^{-\beta E_{x}}} \nonumber \\
		 & = & \exp\big\{\beta F -\beta E_{x}\big\} .
		\label{EqCE}
	\end{eqnarray}

	\noindent In this expression the sum is over all states of the system 
	and
	$F(\beta, {\mathbf E}) =
		-\beta^{-1}\ln \sum_{x}\exp\{-\beta E_{x}\}$ is 
	the Helmholtz free energy of the system.
	
In contrast to an equilibrium ensemble, the probability distribution 
of a nonequilibrium ensemble is not determined solely by the external 
constraints, but explicitly depends on the dynamics and history of the 
system.  Let us consider a stochastic dynamics with a discrete time 
scale, $t\in\{0,1,2,3,\cdots,\tau\}$.  The state of the system at time 
$t$ is $x(t)$, and the path, or trajectory that the system takes 
through this state space can be represented by the vector ${\mathbf 
x}= \bigl( x(0), x(1), x(2),\cdots, x(\tau)\bigr)$.  We make the 
assumption that the dynamics are Markovian\cite{Norris97}.  This 
implies that the probability of 
making a transition between states in a particular time step depends 
only on the current state of the system, and not on the previous 
history.  The single time step dynamics are determined by the 
transition matrix $M(t)$ whose elements are the transition 
probabilities;
	\begin{equation} 
	M(t)_{x{\scriptscriptstyle (t+1)} x{\scriptscriptstyle (t)}} 
		\equiv { P}\bigl(x(t) \rightarrow x(t\sadd1)\bigr) \, .
	\end{equation}
	\noindent A transition matrix, $M$, has the properties that all elements must be 
	nonnegative and that all columns sum to $1$ due to the normalization of 
	probabilities:
	\begin{eqnarray*}
		M _{i j}&\geq&0 \qquad \mbox{ for all } i \mbox{ and } j , \\
		\sum_{i} M_{i j}&=&1	\qquad \mbox{ for all } j .
	\end{eqnarray*}

	Let ${\mathbf \rho}(t)$ be a (column) vector whose elements are the 
	probability of being  in state $i$ at time $t$. Then the single time 
	step dynamics can be written as
	\begin{equation}
	 	\rho(t\sadd1) =  M(t)\,\rho(t) ,
	 	\label{EqFTrans}
	\end{equation}
	\noindent or equivalently as
	\begin{equation}
		\rho(t\sadd1)_{i} = 
		\sum_{j} M(t)_{ij} \,\rho(t)_{j} \,.
	\nonumber
	\end{equation}

	The state energies ${\mathbf 
	E}(t)$ and the
	transition matrices $M(t)$ 
	are  functions of time  due to the external perturbation of the 
	system, and the resulting Markov chain is
	non-homogeneous in time\cite{Vassiliou97}.  The vector of transition matrices 
	${\mathbf M} = \bigl(M(0), M(2), \cdots, M(\tau\ssub1) \bigr)$  completely 
	determine the dynamics of the system.
	We place the following additional constraints on the dynamics; that 
	the 
	state energies are always finite (this avoids the possibility of an 
	infinite amount of energy being transferred from or to the system),
	 and that the single time step transition matrices
	must preserve the corresponding canonical distribution.  This 
	canonical distribution, Eq.~(\ref{EqCE}), is determined by 
	the temperature of the heat bath and the state energies at that time 
	step. We say that the transition matrix is balanced, or that the 
	equilibrium distribution $\pi(t)$
	 is an invariant distribution of $M(t)$.
	$$	\pi(t) =  M(t)	\, \pi(t) $$
	
	\noindent Essentially this condition says that if the system is 
	already in equilibrium (given ${\mathbf E}(t)$ and $\beta$), and the 
	system is unperturbed, then it must remain in equilibrium. 

	It is often convenient to impose the 
	much more restrictive condition of detailed balance, 
	\begin{equation}
		 M(t)_{ij} \,\pi(t)_{j} = M(t)_{ji}\, \pi(t)_{i}  \,.
		\label{EqDB}
	\end{equation}
		
	\noindent In particular many Monte-Carlo simulations are detailed 
	balanced. However, it is not a necessity in such simulations\cite{Man98},
    and it is not necessary here. It is sufficient that the transition 
    matrices are balanced.

Each time step of this dynamics can be separated into two distinct substeps.	
At time $t=0$ the system is in state $x(0)$ with energy 
$E(0)_{x\!\!\:{\scriptscriptstyle (0)}}$.  In the first substep the 
system makes a stochastic transition to a state $x(1)$ which has 
energy $E(0)_{x\!\!\:{\scriptscriptstyle (1)}}$.  This causes an 
amount of energy, $E(0)_{x\!\!\:{\scriptscriptstyle 
(1)}}-E(0)_{x\!\!\:{\scriptscriptstyle (0)}}$, to enter the system in 
the form of heat.  In the second substep the state energies change 
from ${\mathbf E}(0)$ to ${\mathbf E}(1)$ due to the external 
perturbation acting on the system.  This requires an amount of work, 
$E(1)_{x\!\!\:{\scriptscriptstyle 
(1)}}-E(0)_{x\!\!\:{\scriptscriptstyle (1)}}$.  This sequence of 
substeps repeats for a total of $\tau$ time steps.  The total heat 
exchanged with the reservoir, $\mathcal Q$, the total work performed 
on the system, $\mathcal W$, and the total change in energy, $\Delta 
E$, are therefore %
	\begin{eqnarray}
		{\mathcal Q}[{\mathbf x}] &=&
			\sum^{\tau-1}_{t=0}
			\Big[E(t)_{x\!\!\:{\scriptscriptstyle (t+1)}}-E(t)_{x\!\!\:{\scriptscriptstyle 
			(t)}}\Big],
			\label{EqHeat}
			\\
		{\mathcal W}[{\mathbf x}] &=&
			\sum^{\tau-1}_{t=0}\Bigl[
			E(t\sadd1)_{x\!\!\:{\scriptscriptstyle (t+1)}}-E(t)_{x\!\!\:{\scriptscriptstyle 
			(t+1)}}\Bigr],
			 \\[0.1in]
			 	\Delta E &=& E(\tau)_{x\!\!\:{\scriptscriptstyle (\tau)}}-E(0)_{x\!\!\:{\scriptscriptstyle (0)}}
				= {\mathcal W} + {\mathcal Q}. 
	\end{eqnarray}
	
	\noindent The reversible work, 
	$W_{\mathrm r} = \Delta F = F\bigl(\beta,{\mathbf E}(\tau) \bigr) -F\bigl(\beta,{\mathbf 
	E}(0)\bigr)$,
	is the free energy difference between two equilibrium ensembles.  
	It is the minimum average amount of work required to change one 
	ensemble into another.  The dissipative work, ${\mathcal 
	W}_{\mathrm d}[{\mathbf x}] ={\mathcal W}[{\mathbf x}] - W_{r}$, is defined 
	as the difference between the actual work and the reversible work.  
	Note that the total work, the dissipative work and the heat are 
	all path functions.  In this paper they are written with script letters, square 
	brackets and/or as functions of the path, $\mathbf x$, to 
	emphasize this fact.  In contrast $\Delta E$ is a state function;
	it depends only on the initial and final state.

	Now we will consider the effects of a time reversal on this Markov chain.
	In many contexts a time reversal is implemented by  permuting the 
	states of the system. For example, in a Hamiltonian system a time 
	reversal involves inverting the momenta of all the particles. 
	However, it is equivalent, and in the current context much more 
	convenient, to apply the effects of the time reversal to the dynamics 
	rather than the state space. Thus the time-reversed trajectory, 
	$\widehat{\mathbf x}$, is a simple reordering of the forward 
	trajectory; $\widehat{x}(t) =x(\tau\ssub t)$ and 
	$\widehat{\mathbf E}(t) ={\mathbf E}(\tau\ssub t)$.
	
	We can derive the effect of a time reversal on a transition matrix by 
	considering a time homogeneous Markov chain.  Let $\pi$ be the invariant 
	distribution of the time-independent transition matrix $M$, given by the
	 equilibrium canonical  ensemble.  If the system is in an 
	 equilibrium ensemble then a time reversal should 
	have no effect on that ensemble, and the probability of observing 
	the transition $i\rightarrow j$ in the forward chain should be the 
	same as the probability of observing the transition $j\rightarrow i$ 
	in the time-reversed chain. Because the equilibrium probability distribution
	 is the same for both chains it follows that
	\begin{equation}
		 \widehat{M}_{ji} \,\pi_{i}=  M_{ij} \,\pi_{j}\qquad \mbox{for all} \quad i,j
	\,.
	\end{equation}
	\noindent In matrix notation this may conveniently be written as 
	$$	\widehat{M} = \mbox{diag}(\pi)^{-1} M^{T} \mbox{diag}(\pi) .$$
	
	\noindent Here, $\mbox{diag}(\pi)$ indicates a matrix whose diagonal 
	elements are given by the vector $\pi$. $\widehat{M}$ is referred to as 
	the reversal of $M$\cite{Norris97}, or as 
	the $\pi$-dual of $M$\cite{KSK76}.  If the transition matrix obeys detailed 
	balance, Eq.~(\ref{EqDB}), then $\widehat{M}=M$.
	
	It is easy to confirm that $\widehat{M}$ is a transition matrix; all entries are 
	nonnegative because all equilibrium and transition probabilities are 
	nonnegative, and all rows sum to $1$,
	$$ 
		\sum_{j} \widehat{M}_{ji} = \frac{1}{\pi_{i}} \sum_{j}  M_{ij} \,\pi_{j} 
		=\frac{\pi_{i}}{\pi_{i}} =1 \qquad \mbox{for all} \quad i \, .
	$$
	Further, we can demonstrate that $\widehat{M}$ and $M$ have the same invariant distribution,
	$$ 
		\sum_{i} \widehat{M}_{ji}\, \pi_{i}  =  \sum_{i}  M_{ij} \,\pi_{j}
		=\pi_{j} \, .
	$$
	
	For the non-homogeneous chain the time reversal of the vector of transition 
	matrices, $\mathbf M$, is defined as
	\begin{equation}
			\widehat{M}(t) = \mbox{diag}(\pi(\tau\ssub t))^{-1} 
		M(\tau\ssub t)^{T}\mbox{diag}(\pi(\tau\ssub t))
	.\label{EqTR}
	\end{equation}
	\noindent 
	The 
	time reversal operation is applied to each transition matrix, and 
	their time order is reversed.  Note that for the transition 
	matrices of the reverse chain the time index runs from $1$ to 
	$\tau$, rather than $0$ to $\tau\ssub1$. Therefore,  $M(t)$ is the transition matrix 
	from time 
	$t$ to time $t\sadd1$ (see Eq.~(\ref{EqFTrans})),  but 
	$\widehat{M}(t)$ is the transition matrix  from 
	time $t\ssub1$ to time $t$.
	\begin{equation}
		\widehat{\rho}(t) =
		 \widehat{M}(t) \,\widehat{\rho}(t\ssub1)  \, .
		\label{EqRTrans}
	\end{equation}	
	This convention is 
	chosen so that the time indexes of the various entities remains 
	consistent.  Thus for the reverse chain at time $t$ the state  
	is $\widehat{x}(t)$, the states energies are ${\widehat{\mathbf E}}(t)$ and the corresponding 
	equilibrium distribution is $\widehat{\pi}(t)$, which is an invariant 
	distribution of $\widehat{M}(t)$. 
	
	Another consequence of the time reversal is that the work and heat 
	substeps are interchanged in the 
	reverse chain. The heat, total work and dissipative work are all odd 
	under a time reversal:
	${\mathcal Q}[{\mathbf x}]=-{\mathcal Q}[\widehat{\mathbf x}]$,
	 ${\mathcal W}[{\mathbf x}]=-{\mathcal W}[\widehat{\mathbf x}]$
 	and
	${\mathcal W}_{\mathrm d}[{\mathbf x}]=-{\mathcal W}_{\mathrm d}[\widehat{\mathbf x}]$. The total change in energy, and 
	the free energy change are also odd under a time reversal, but to 
	avoid ambiguity a `$\Delta$' always refers to a change measured 
	along the forward process.

	We are now in a position to prove an important symmetry for the driven 
	system under consideration.  Let ${\mathcal P}[\,{\mathbf x}\,|\, x(0),{\mathbf M}\, ]$
	be the probability of the trajectory $\mathbf x$, given that the 
	system 
	started in state $x(0)$. The  probability of the 
	corresponding reversed path is $\widehat{\mathcal P}[\,{\widehat{\mathbf 
	x}}\,|\,\widehat{x}(0),\widehat{\mathbf M}\,]$. The ratio of these path 
	probabilities is a simple function of the heat exchanged with the 
	bath,
	 \begin{equation}
	 	\frac{ {\mathcal P}[\,{\mathbf x}\,|\, x(0),{\mathbf M}\, ]}
	 		 { \widehat{\mathcal P}[\,{\widehat{\mathbf x}}\,|\,\widehat{x}(0),\widehat{\mathbf 
	 		 M}\,]}
	 		 	 = \exp\big\{\ssub\beta {\mathcal Q}[{\mathbf x}]\big\}
	 		 	.
      		\label{EqMR}
	\end{equation}

	\noindent At the risk of ambiguity, a system with this property will be 
	described as microscopically 
	reversible\cite{Crooks98a,Crooks99b,MaesRedigMoffaert99}.

	We proceed by expanding the 
	path probability as a product of single time step transition probabilities.  
	This follows from the condition that the dynamics are Markovian.
	$$
	\frac{{\mathcal P}[\,{\mathbf x}\,|\,{x}(0),\mathbf M\,]} 
		{\widehat{\mathcal P}[\,{\widehat{\mathbf x}}\,|\,\widehat{x}(0),\widehat{\mathbf M}\,]} =
	 \frac{\displaystyle\prod_{t=0}^{\tau-1}
	 			{ P}\bigl(x(t)\rightarrow  x(t\sadd1)\bigr)
	 			\hphantom{{}'{}'}
	 			}
			{\displaystyle \prod_{t'=0}^{\tau-1}
				\widehat{ P}\bigl(\widehat{x}(t')\rightarrow 
							\widehat{x}(t'\sadd1)\bigr)}	
	$$

	For every transition in the forward chain there is a transition in 
	the reverse chain related by the time reversal symmetry, Eq.~(\ref{EqTR}). 
	 The path probability ratio can therefore be converted into a product of 
	equilibrium probabilities.
	\begin{eqnarray*}
		\frac{{\mathcal P}{[\,{\mathbf x}\,|\,\widehat{x}(0),{\mathbf M}\,] }} 
			{\widehat{\mathcal P}[\,\widehat{\mathbf x}\,|\,\widehat{x}(0),\widehat{\mathbf 
			M}\,] } &=&
		\prod_{t=0}^{\tau-1}  \frac{\pi(t)_{x\!\!\:\scriptscriptstyle (t+1)}}
								{\pi(t)_{x\!\!\:\scriptscriptstyle (t)\hphantom{+1}}} 
		=\prod_{t=0}^{\tau-1}  
			\frac{\rho(x(t\sadd1)|\beta,{\mathbf E}(t))}
				{\rho(x(t)|\beta,{\mathbf E}(t))} 
						\\
		&=&\exp \Bigl\{\ssub\beta\sum_{t=0}^{\tau-1} 
		\bigl[ E(t)_{x\!\!\:\scriptscriptstyle (t+1)} -E(t)_{x\!\!\:\scriptscriptstyle (t)} 
		\bigr]
		\Bigr\} \\[0.125in]
		&=& \exp\big\{\ssub\beta {\mathcal Q}[{\mathbf x}]\big\}
	\end{eqnarray*}
	\noindent The second line follows from the definition of the 
	canonical ensemble, Eq.~(\ref{EqCE}), and the final line from the 
	definition of the heat, Eq.~(\ref{EqHeat}).

	The essential assumptions leading to this condition of 
	microscopic reversibility are that the state energies are always finite, 
	and that the dynamics are Markovian, and if unperturbed preserve 
	the equilibrium distribution. These conditions are valid independently 
	of the strength of the perturbation, or the distance of the ensemble from equilibrium.
	The extension to continuous time and continuous phase-space appears 
	straightforward, although it is technically more difficult to be 
	completely rigorous.  However, Jarzynski\cite{Jarzynski99b} has recently 
	demonstrated that deterministic Hamiltonian system coupled to many 
	heat baths are also microscopically reversible.

\section{Path ensemble averages}
	
	We are now in a position to consider the path ensemble average 
	(Eq.~(\ref{EqPathEnsembleAvg})) detailed in the introduction. A system that is initially in 
	thermal equilibrium is  driven  away from that equilibrium by an 
	external perturbation, and the path function
	 ${\mathcal F}[ \mathbf{x} ]$ is averaged over the resulting 
	 nonequilibrium ensemble of paths. The probability of a 
	trajectory is determined by the equilibrium probability of the 
	initial state, and by the vector of 
	transition matrices that determine the dynamics. Therefore, the average 
	of $\mathcal F$ over the ensemble of trajectories can be explicitly 
	written as
	$$	\big\langle {\mathcal F}\big\rangle_{\!\mathrm F} = 
					\sum_{ \mathbf x  }
					\rho\bigl(x(\tb 0\tb )
						\big|\beta, {\mathbf E}(\tb 0\tb )\bigr) \,
					 {\mathcal P}[\,{\mathbf x}\,|\,x(0),{\mathbf M}]
						\,{\mathcal F}[\mathbf x] \, .
	$$

	\noindent	The sum is over the set of all paths connecting all 
	possible initial and final states. Given that 
	the system is microscopically reversible it is a simple 
	matter to convert the above expression to an average over the 
	reverse process. We first note that
	\begin{eqnarray}
		\frac{\rho\bigl(x(\tb 0\tb )
						\big|\beta, {\mathbf E}(\tb 0\tb )\bigr)  \,
					{\mathcal P}[{\mathbf x}|{x}(0),{\mathbf M}]} 
			{\rho\bigl(\widehat{x}(\tb 0\tb )
						\big|\beta, \widehat{\mathbf E}(\tb 0\tb )\bigr)  \,
				\widehat{\mathcal P}[\widehat{\mathbf x}|\widehat{x}(0),\widehat{\mathbf M}]}
		&=& e^{+\beta \Delta E \ssub\beta \Delta F \ssub\beta {\mathcal Q}[{\mathbf x}]}, \nonumber \\
		&=& e^{+\beta {\mathcal W}[{\mathbf x}] \ssub\beta \Delta 
		F} ,\nonumber \\[0.125in]
		&=& e^{+\beta {\mathcal W}_{\!\mathrm d}[{\mathbf x}]} .
	\end{eqnarray}
	
	\noindent The first line follows from the condition of microscopic reversibility Eq.~(\ref{EqMR}), and the definition of the 
	canonical ensemble, Eq.~(\ref{EqCE}). Recall that $\Delta F$ is the reversible 
	work of the forward process, 
	and that ${\mathcal W}_{\!\mathrm d}[x]$ is the dissipative work.
	The set of reverse trajectories is the 
	same as the set of forward trajectories, and we define 
	${\mathcal F}[{\mathbf x}] = \widehat{\mathcal F}[\widehat{\mathbf x}]$.

	Therefore,
	\begin{eqnarray*}
		\big\langle {\mathcal F}\big\rangle_{\!\mathrm F} &=& 
					\sum_{ \mathbf {\displaystyle \widehat{\mathbf x}} } 
					\rho\bigl(\widehat{x}(\tb 0\tb )
						\big|\beta, \widehat{\mathbf E}(\tb 0\tb )\bigr) \,
					 \widehat{\mathcal P}[\,{\widehat{\mathbf x}}\,|\,\widehat{\mathrm M}\,]
						\,{\mathcal \widehat{F}}[{\mathbf \widehat{x}}] \,\,e^{-\beta 
						{\mathcal W}_{\!\mathrm d}{\displaystyle [\hat{\mathbf x}]} }\\
		&=& \big\langle \widehat{\mathcal F} \,e^{-\beta {\mathcal W}_{\!\mathrm d}} \big\rangle_{\!\mathrm R}
		\, .			
	\end{eqnarray*}
 
 	It is frequently convenient to  rewrite Eq.~(\ref{EqPathEnsembleAvg}) as
	\begin{equation}
		\big\langle  {\mathcal F}\,e^{-\beta {\mathcal W}_{\!\mathrm d}}\big\rangle_{\!\mathrm F}
			=\big\langle \widehat{\mathcal F}  \big\rangle_{\!\mathrm R} ,
	\label{EqPathEnsembleAvg2}
	\end{equation}
	
	\noindent where  $\mathcal F$ has been  replaced with ${\mathcal F}\,e^{-\beta 
	{\mathcal W}_{\!\mathrm d}} $, and $\widehat{\mathcal F}$ with 
	$\widehat{\mathcal F}\,e^{+\beta {\mathcal W}_{\!\mathrm d}}$.

\subsection{Jarzynski nonequilibrium work relations}

A variety of previous known relations can be considered special cases 
or approximations of this nonequilibrium path ensemble average.  In 
the simplest case we start with Eq.~(\ref{EqPathEnsembleAvg2}), and 
then set ${\mathcal F} = \widehat{\mathcal F} = 1$ (or any other
constant of the dynamics).  Then %
	\begin{equation}
		\big\langle  e^{-\beta {\mathcal W}_{\!\mathrm d}} \big\rangle_{\!\mathrm F}
			=\big\langle 1 \big\rangle_{\!\mathrm R} =1
	.
	\end{equation}

	\noindent The right side is unity due to normalization of probability distributions. 
	We are now taking an average over a single path ensemble, and the 
	remaining subscript, ``F'', becomes superfluous.  The dissipative 
	work, ${\mathcal W}_{\!\mathrm d}$ can replaced by ${\mathcal W}-\Delta 
	F$, and the change in free energy can be moved outside the average 
	since it is path independent.
	 The result is the Jarzynski nonequilibrium work 
	relation\cite{Jarzynski97a,Jarzynski97b,Jarzynski98b,Crooks98a,Crooks99b}.
	\begin{equation}
		 \langle e^{-\beta W }\rangle =e^{-\beta \Delta F}
		\label{EqJarzynski}
	\end{equation}
	
	\noindent This relation states that if we convert one system into 
	another by changing the energies of all the states from an initial 
	set of values to a final set of values over some finite length of 
	time, then the change in the free energies of the corresponding equilibrium 
	ensembles can be calculated by 
	repeating the switching process many times, each time starting from 
	an equilibrium ensemble, and taking the above average of the amount 
	of work required to effect the change.
	 In the limit of instantaneous switching between ensembles, (we change 
	 the energies of all the states in a single instantaneous jump) this relation is 
	 equivalent to the standard thermodynamic perturbation method that 
	 is frequently used to calculate free energy differences by computer 
	 simulation\cite{Frenkel96}.

It is possible to extend Eq.~(\ref{EqJarzynski}) to a more general 
class of relations between the work and the free energy 
change\cite{Jarzynski98p}.  Suppose that ${\mathcal F} = f(\mathcal 
W)$ where $f({\mathcal W})$ is any finite function of the work.  Then 
$\widehat{\mathcal F} = f({\ssub\mathcal W})$, because the work is odd 
under a time reversal.  Inserting these definitions into 
Eq.~(\ref{EqPathEnsembleAvg}) and rearranging gives %
\begin{equation}
	e^{-\beta \Delta F } = \frac{\bigl\langle  f(\sadd{\mathcal W}) \bigr\rangle_{\!\mathrm F} }{ 
	\bigl\langle f(\ssub{\mathcal W}) \,e^{-\beta {\mathcal W}}\bigr\rangle_{\!\mathrm R}}
\end{equation}

Recall that $\Delta F$ is defined in terms of the forward process.  
Suppose that we have obtained $n_{\scriptscriptstyle \mathrm F}$ 
independent measurements of the work required for the forward process, 
and $n_{\scriptscriptstyle \mathrm R}$ independent measurements from 
the reverse process.  An interesting question is what choice of $f({ 
\mathcal W})$ leads to the highest statistical accuracy for $\Delta 
F$.  For instantaneous switching this question was answered by 
Bennett\cite{Bennett76,Frenkel96} in his derivation of the acceptance 
ratio method for calculating free energy differences.  For finite time 
switching Bennett's derivation can be followed almost line for line.  
We therefore omit the details, and simply record the conclusions in 
the present notation.  The least statistical error will result if we 
take ${\mathcal F}=(1\sadd\exp\{\sadd\beta {\mathcal W} \sadd 
C\})^{-1}$, and $\widehat{\mathcal F}=(1\sadd\exp\{\ssub\beta 
{\mathcal W} \sadd C\})^{-1}$.  Then 
	\begin{equation}
		e^{-\beta \Delta F } = \frac{\bigl\langle  (1\sadd\exp\{\sadd\beta {\mathcal 
		W} \sadd C\})^{-1} \bigr\rangle_{\!\mathrm F} }
				{ \bigl\langle (1\sadd\exp\{\sadd\beta {\mathcal W} 
				\ssub C\})^{-1}\bigr\rangle_{\!\mathrm R}}
				\,\exp\{\sadd C\} \,.
	\end{equation}
	
	\noindent The optimal choice of the constant $C$ is $\ssub\beta\Delta F 
	\sadd\ln 
	n_{\scriptscriptstyle \mathrm F}/n_{\scriptscriptstyle \mathrm R}$.
	This relation must be solved self-consistently, since 
	$\Delta F$ appears on both sides.

\subsection{Transient fluctuation theorem}
	Another interesting application of the path ensemble average is to replace the finite 
	function of the work used above with a $\delta$ function, 
	${\mathcal F} =\delta(\beta {\mathcal W}_{\mathrm d} \ssub\beta {\mathcal W}_{\mathrm d}[{\mathbf x}])$, 
	${\mathcal \widehat{F}} =\delta(\beta {\mathcal W}_{\mathrm d} \sadd\beta {\mathcal W}_{\mathrm d}[{\mathbf \widehat{x}}])$. Plugging 
	these $\mathcal F$'s 
	into Eq.~(\ref{EqPathEnsembleAvg}) gives
	\begin{eqnarray*}
		\bigl\langle  \delta(\beta {\mathcal W}_{\mathrm d} \ssub\beta {\mathcal W}_{\mathrm d}[{\mathbf 
		x}])\,\,e^{-\beta {\mathcal W}_{\!\mathrm d}} \bigr\rangle_{\!\mathrm F}
			&=&\bigl\langle \delta(\beta {\mathcal W}_{\mathrm d} \sadd\beta {\mathcal 
			W}_{\mathrm d}[{\mathbf  \widehat{x}}])   \bigr\rangle_{\!\mathrm R} 
			\,\, , \\[8pt]
			\mbox{or}\qquad\qquad	
				{ P_{\!\!\scriptscriptstyle \mathrm F}}(\sadd\beta {\mathcal W}_{\mathrm d}) \, e^{-\beta {\mathcal W}_{\mathrm d}} 
				&=& { P_{\!\!\scriptscriptstyle \mathrm R}}(\ssub\beta {\mathcal W}_{\mathrm 
				d}) \,.
	\end{eqnarray*}

	\noindent Here, ${ P_{\!\!\scriptscriptstyle \mathrm F}}(\sadd\beta {\mathcal W}_{\mathrm d})$ is the 
	probability of expending the specified amount of work in the forward 
	process, and ${ P_{\!\!\scriptscriptstyle \mathrm R}}(\ssub\beta {\mathcal W}_{\mathrm d})$ is the 
	probability of expending the negative of that amount of work in the 
	reverse process. If ${ P_{\!\!\scriptscriptstyle \mathrm R}}(\ssub\beta {\mathcal W}_{\mathrm d}) 
	\neq 0$ then we can 
	rearrange this expression as
	\begin{equation}
		\frac{ P_{\!\!\scriptscriptstyle \mathrm F}(\sadd\beta {\mathcal 
		W}_{\mathrm d}) }
		{P_{\!\!\scriptscriptstyle\mathrm R}(\ssub\beta 
		{\mathcal W}_{\mathrm d})}  = e^{+\beta {\mathcal W}_{\mathrm d}} .
			\label{EqTransFT}
	\end{equation}
	
	The system of interest starts in equilibrium and is perturbed for a finite 
	amount of time. If it is allowed to relax back to equilibrium then 
	the change in entropy of the heat bath will be $\ssub\beta {\mathcal Q}$, 
	and the change in entropy of the system will be $\beta\Delta E\ssub\beta\Delta F$. 
	Therefore, the total change in entropy of the universe resulting from 
	the perturbation of the system is $\ssub\beta {\mathcal Q} \sadd\beta\Delta 
	E\ssub\beta\Delta F = \beta {\mathcal W} \ssub\beta\Delta F= \beta{\mathcal W}_{\!\mathrm d}$, the dissipative work. Thus 
	Eq.~(\ref{EqTransFT}) can be interpreted as an entropy production fluctuation 
	theorem. It relates the  distribution of entropy productions of
	a driven system that is initially in equilibrium to the entropy 
	production of the same system driven in reverse. As such it is closely 
	related to the transient fluctuation theorems of Evans and 
	Searles\cite{EvansSearles94,EvansSearles96}.  The connection between 
	this fluctuation theorem, the Jarzynski nonequilibrium work 
	relation and microscopic reversibility was originally presented in 
	\cite{Crooks99b}.

	\subsection{Kawasaki response and nonequilibrium distributions}
	
	All of the above relations were derived from Eq.~(\ref{EqPathEnsembleAvg}) by inserting a
	function of the work. Another group of relations can be derived by 
	instead setting $\mathcal F$ to be a function of the state of the 
	system at some time. In particular if we average a function of the 
	final state in the forward process, ${\mathcal F} 
	=f\bigl(x(\tb \tau\tb )\bigr)$, then we average a function of the initial state in the 
	reverse process, $\widehat{\mathcal 
	F}=f\bigl(\widehat{x}(\tb 0\tb )\bigr)$:
	$$
		\bigl\langle  f\bigl(x(\tb \tau\tb )\bigr)\,e^{-\beta {\mathcal 
		W}_{\!\mathrm d}} \bigr\rangle_{\!\mathrm F}
			=\bigl\langle f\bigl(\widehat{x}(\tb 0\tb )\bigr) \bigr\rangle_{\!\mathrm R}
			\, . 
	$$
	
	\noindent 	 Therefore, in the reverse process
	 the  average is over the initial 
	 equilibrium ensemble of the system, and the subsequent dynamics are 
	 irrelevant. We can once more drop 
	reference to forward or reverse processes, and instead  use 
	labels to indicate equilibrium and nonequilibrium averages:
	\begin{equation}
		\bigl\langle  f\bigl(x(\tb \tau\tb )\bigr)\,e^{-\beta {\mathcal 
		W}_{\!\mathrm d}} \bigr\rangle_{\!\mathrm neq}
			=\bigl\langle f\bigl(x(\tb \tau\tb )\bigr) 
			\bigr\rangle_{\!\mathrm eq} \, .
	\end{equation}

	\noindent 
	This relation (also due to Jarzynski\cite{Jarzynski98p}) states that the average of a state function 
	in a nonequilibrium ensemble,
	 weighted by the dissipative work, can be equated with 
	an equilibrium average of the same quantity.
	
	Another interesting relation results if we insert the same state 
	functions into the alternative form of the path ensemble average, 
	Eq.~(\ref{EqPathEnsembleAvg2}): (This is ultimately equivalent to switching 
	$\mathcal F$ and $\widehat{\mathcal F}$.)
	\begin{equation}
		\bigl\langle  f\bigl(x(\tb \tau\tb )\bigr) \bigr\rangle_{\!\mathrm F}
			=\bigl\langle f\bigl(\widehat{x}(\tb 0\tb ) 
			\bigr) \,e^{-\beta {\mathcal 
		W}_{\!\mathrm d}}
			\bigr\rangle_{\!\mathrm R} \, .
		\label{EqKawasaki}
	\end{equation}

	This is the Kawasaki nonlinear response 
	relation\cite{YamadaKawasaki67,MorrissEvans85,EvansMorriss90,EvansSearles95,PetravicEvans98}, applied to 
	stochastic dynamics, and generalized to arbitrary forcing.
	This relation can also be written in an explicitly renormalized 
	form\cite{EvansSearles95} 
	by expanding the dissipative work as $-\Delta F +\mathcal W$, and rewriting the free 
	energy change as a work average using the Jarzynski relation, 
	Eq.~(\ref{EqJarzynski}).
	\begin{equation}
		\bigl\langle  f\bigl(x(\tb \tau\tb )\bigr) \bigr\rangle_{\!\mathrm F}
			=\bigl\langle f\bigl(\widehat{x}(\tb 0\tb )\bigr) 
				\,e^{-\beta {\mathcal W}}\bigr\rangle_{\!\mathrm R} 
			\Big/
			\bigl\langle e^{-\beta {\mathcal W}}	\bigr\rangle_{\!\mathrm R}
		\label{EqKawasakiRN}
	\end{equation}
	
	\noindent Simulation data indicates that averages calculated with the 
	 renormalized expression typically have lower statistical 
	 errors\cite{EvansSearles95}.

The probability distribution of a nonequilibrium ensemble can be derived 
from the Kawasaki relation, Eq.~(\ref{EqKawasakiRN}), by setting the 
state function to be ${\mathcal 
F}=f(x(\tau))=\delta\bigl(x-x(\tau)\bigr)$, a $\delta$ function of the 
state of the system at time $\tau$; 
\begin{equation}
	\rho_{\mathrm{neq}}\bigl(x,\tau|{\mathbf M}\bigr) =\rho\bigl(x\big|\beta,{\mathbf 
	E}(\tb \tau\tb )\bigr)
	 \frac{\bigl\langle  e^{-\beta {\mathcal W}}\bigr\rangle_{{\mathrm R},x}}
	 {\bigl\langle e^{-\beta {\mathcal W}}\bigr\rangle_{\!\mathrm 
	 R\hphantom{,x}}}	.
 \label{EqNeqPD}
\end{equation}
	
\noindent Here $\rho_{\mathrm{neq}}\bigl(x,\tau|{\mathbf M}\bigr)$ is the 
nonequilibrium probability distribution and 
$\rho\bigl(x\big|\beta,{\mathbf E}(\tb \tau\tb )\bigr)$ is the 
equilibrium probability of the same state.  The subscript `$x$' 
indicates that the average is over all paths that start in state $x$.  
In contrast the lower average is over all paths starting from an 
equilibrium ensemble.  Thus the nonequilibrium probability of a state 
is, to zeroth order, the equilibrium probability, and the correction 
factor can be related to a nonequilibrium average of the work.

There are several other far-from-equilibrium relations that have been 
derived from, or are related to the Kawasaki response.  The transient 
time correlation function (TTCF)\cite{MorrissEvans87,PetravicEvans97} 
gives another set of relations for the nonlinear response of a system, 
and are reputable of greater practical utility than the Kawasaki 
response relation.  Unfortunately it appears that TTCF can not be 
applied to the systems considered in this paper, since a crucial step 
linking the two formalisms\cite{EvansMorriss90} makes the assumption 
that the dynamics are deterministic, and therefore that only an 
average over initial conditions is needed.  Similarly Evans and 
Morriss have derived several interesting relations for the heat 
capacity of a nonequilibrium steady-state\cite{EvansMorriss90}, but 
again these relations are not generally applicable because it is 
assumed that the probability of a trajectory is independent of the 
temperature of the heat bath.

\section{Conclusions}
All of the relations derived in this paper are directly 
applicable to systems driven far-from-equilibrium.  These relations 
 follow if the dynamics are microscopically reversible in the sense 
of Eq.~(\ref{EqMR}).  This relation was shown to hold if the 
dynamics are Markovian and balanced.  Although I have concentrated on 
stochastic dynamics with discrete time and phase space, this should not 
be taken as a fundamental limitation.  The extension to continuous 
phase space and time appears straightforward, and deterministic 
dynamics can be taken as a special case of stochastic dynamics.

It is a pleasure to thank C.~Jarzynski, D.~Chandler, P.~L.~Geissler and 
T.~McCormick for their valuable discussions and suggestions.  This work was initiated with support from the National Science 
Foundation, under grant No.  CHE-9508336, and completed with support 
from the U.S. Department of Energy under contract No.  
DE-AC03-76SF00098.  


\end{document}